\documentclass[twocolumn,aps,prb]{revtex4} %superscriptaddress,eqsecnum,

\usepackage{amsmath,amssymb,amsfonts,bm}
\usepackage{graphicx,epstopdf}  %epsfig,
\usepackage{color}
\usepackage{extarrows}
\usepackage{enumitem}
\usepackage{empheq}
\usepackage{chngcntr}

\renewcommand{\d}{{\rm d}}

\newcommand{\cf}{cf.\ }

\newcommand{\w}{\omega}

\newcommand{\ti}{\tilde}

\newcommand{\E}{\mbox{\tiny E}}

\newcommand{\tS}{\mbox{\tiny S}}
\newcommand{\T}{\mbox{\tiny T}}
\newcommand{\ET}{\mbox{\tiny ET}}

\newcommand{\la}{\langle}
\newcommand{\ra}{\rangle}

\newcommand{\La}{\big\la}
\newcommand{\Ra}{\big\ra}

\newcommand{\Sec}[1]{Sec.\,\ref{#1}}

\newcommand{\nl}{\nonumber \\}
\newcommand{\be}{\begin{equation}}
\newcommand{\ee}{\end{equation}}
\newcommand{\bsube}{\begin{subequations}}
\newcommand{\esube}{\end{subequations}}
\newcommand{\Eq}[1]{Eq.\,(\ref{#1})}
\newcommand{\Eqs}[1]{Eqs.\,(\ref{#1})}
\newcommand{\Fig}[1]{Fig.\,\ref{#1}}

\newcommand{\RN}[1]{%
  \textup{\uppercase\expandafter{\romannumeral#1}}%
}

\allowdisplaybreaks[1]

\makeatletter
\newcommand{\oset}[3][0ex]{%
  \mathrel{\mathop{#3}\limits^{
    \vbox to#1{\kern-2\ex@
    \hbox{$\scriptstyle#2$}\vss}}}}
\makeatother

\begin{document}

\title{Marcus' electron transfer rate revisited
 via a Rice--Ramsperger--Kassel--Marcus analogue: A unified formalism for linear and nonlinear solvation scenarios}
%%%

\author{Yao Wang}
\email{wy2010@ustc.edu.cn}
\author{Yu Su}
\author{Rui-Xue Xu}%\email{rxxu@ustc.edu.cn}
\author{Xiao Zheng}%\email{xz58@ustc.edu.cn}
\author{YiJing Yan}%\email{yanyj@ustc.edu.cn}

\affiliation{Hefei National Laboratory for Physical Sciences at the Microscale
and Department of Chemical Physics
and Synergetic Innovation Center of Quantum Information and Quantum Physics
and
Collaborative Innovation Center of Chemistry for Energy Materials (i{\rm ChEM}),
University of Science and Technology of China, Hefei, Anhui 230026, China}

\date{January 5, 2021}

\begin{abstract}

In the pioneering work by R. A. Marcus,
the solvation effect on electron transfer (ET) processes
was investigated, giving rise to the celebrated nonadiabatic ET rate formula.
%%%
In this work, on the basis of the thermodynamic solvation potentials analysis, we reexamine Marcus' formula with respect to
the Rice--Ramsperger--Kassel--Marcus (RRKM) theory.
Interestingly, the obtained RRKM analogue,
which recovers the original Marcus' rate that is in a linear solvation
scenario, is also applicable to the nonlinear solvation scenarios, where the
multiple curve--crossing of solvation potentials exists.
%%%
Parallelly, we revisit the corresponding Fermi's golden rule results, with some critical comments against
the RRKM analogue proposed in this work. For illustration, we consider the
quadratic solvation scenarios,
on the basis of physically
well--supported descriptors.

\end{abstract}
\maketitle

\section{Introduction}
 Electron transfer (ET) is a fundamental and representative type of  physical--chemistry processes.
The Marcus' ET theory is considered to be a milestone for understanding the solvation effect in these processes.
%%%
It gives rise to the celebrated nonadiabatic ET rate,
which reads\cite{Mar56966,Mar64155,Sum864894,Mar93599}
\be\label{ET_rate}
 k = \frac{V^2/\hbar}{\sqrt{\lambda k_{B}T/\pi}}
  \exp\bigg[-\frac{(E^{\circ}+\lambda)^2}{4\lambda k_{B}T}\bigg].
\ee
Here, $V$ denotes the nonadiabatic transfer coupling parameter
and $E^{\circ} \simeq \Delta G^{\circ}_r$
%\simeq \Delta G^{\mbox{\tiny$\ominus$}}_r$
amounts to
the standard reaction Gibbs energy,
for the electron transferring
from the donor ($|0\ra$; reactant) to the acceptor ($|1\ra$; product) state,
with $\lambda$ being the associated solvent reorganization energy.

The minimum model for the total ET composite Hamiltonian assumes
\be\label{HT0}
  H_{\ET}=h_{0}|0\ra\la 0|+(E^{\circ}+h_{1})|1\ra\la 1|
   +V\big(|0\ra\la 1|+|1\ra\la 0|\big).
\ee
The electronic system in donor and acceptor
states are associated with their own solvent environments.
The individual solvent environment is characterized by
not only the Hamiltonian, $h_{a}$, but also
 $\rho^{\E;{\rm eq}}_{a}(T)
 \equiv e^{-\beta h_{a}}/{\rm tr}e^{-\beta h_{a}}$,
with $\beta=1/(k_BT)$, where $k_{B}$ is the Boltzmann constant
and $T$ the temperature.
The total ET composite was initially
$\rho_{\T}(t_0)=\rho^{\E;{\rm eq}}_{0}(T)|0\ra\la 0|$,
the thermal equilibrium in the donor state,
prior to the nonadiabatic coupling $V$ taking action.
%%%

It is widely known that the Marcus' nonadiabatic ET rate formula,
\Eq{ET_rate}, can be derived via the Fermi's golden rule (FGR) in the linear solvation scenarios (\cf\,\Sec{thsec2A}). 
However, in reality, some 
degrees of freedom that can be
treated as a part of solvent environment
are nonlinear.\cite{Pen079333,Wan0710369,Zha121075,Zan1613351,%
Yan19074106,Hsi18014104,Hsi20125002,Hsi20125003}
Can the FGR approach be directly extended to nonlinear solvation scenarios? And why is this the case? To answer these questions, other perceptions of understanding Marcus' ET rate formula may be helpful.

To this end, it would be interesting to see whether the Marcus' ET rate formula
can be reproduced
from the prospect of Rice--Ramsperger--Kassel--Marcus (RRKM)
theory.\cite{Ric271617,Ric28617,Kas28225,Kas281065,Mar51894,Mar52359,Hol96}
%%%
This theory shares the common ground with
the transition--state theory (TST).\cite{Eva35875,Eyr35107,Eyr3565}
They both deal with the adiabatic events of first--passage
over the barrier top where the activated complex is located.
%%%%
The underlying reaction kinetic mechanism is described with
\be\label{TST_mechanism}
 {\rm R} \overset{\text{pre-eq.}}
\rightleftharpoons  {\rm X}^{\ddag}
%%
%\overset{k^{\ddag}}
%{\overset{\text{slow}}
%%
%{\longrightarrow}} 
\oset[1.0ex]{\text{slow}}{\longrightarrow}{\rm P}
\ee
It starts with the fast pre-equilibrium step, forming
activated complex X$^{\ddag}$, and then undergoes a slow
``unimolecular dissociation'' to the product (P).
%%%
For the latter step, from the activated complex to product,
the TST describes with the Eyring's reactive
coordinate frequency formula,\cite{Eyr35107,Eyr3565}
whereas the RRKM theory goes by an ergodicity description,
adopted originally by Hinshelwood.\cite{Hin26}
%%%
Nevertheless, both TST and RRKM theory are considered
to be the methods for adiabatic gas--phase rate processes.
%%%
Can the RRKM be applied to nonadiabatic condensed--phase ET rate processes? 

In both TST and RRKM theory,
the reactant (R) involved in the first step of
kinetic mechanism (\ref{TST_mechanism}) consists of two gas molecules.
%%%
What will be their correspondences
in the condensed--phase ET rate processes?
{\color{black}
This issue could be addressed on the basis of
the statistical quasi-particle description for the
solvation effect on ET
processes.\cite{Yan14054105,Yan16110306,Zha18780,Wan20041102}
%%%
In other words, we visualize the solvation
as a collective species. 
}
It collides with a solute molecule,
forming an activated solute--solvent complex X$^{\ddag}$.
%%%
This first step of the kinetic mechanism (\ref{TST_mechanism})
is a fast pre-equilibrium process.
The second step assumes a slow solvent relaxation,
resulting in a stable ET product.

Based on the above ideas, in this work we propose a RRKM analogue
that not only recovers the Marcus' nonadiabatic ET rate,
\Eq{ET_rate}, in the linear solvation scenario (\cf\,\Sec{theSec2B}), but also treats
the nonlinear solvation scenarios, where the
multiple curve--crossing of solvation potentials exists (\cf\,\Sec{thesec3B}). Parallelly, we revisit the corresponding FGR results for comparisons (\cf\,\Sec{thesec3A}).

As an illustration, we consider the
quadratic solvation scenarios,
on the basis of physically
well--supported descriptors (\cf\,\Sec{thsec4}).\cite{Xu17395,Xu18114103,Liu18245}
{\color{black}
The results are discussed in different cases,
with heuristic theoretical pictures.
In \Sec{thsum} are concluding remarks, 
together with the prospect of this work.
}

\section{Linear solvation scenario}\label{thsec2}
In the linear solvation scenarios, the solvation coordinate ($X$) is
\be\label{U_def}
  X=U \equiv h_{1}-h_{0}.
\ee
It is worth emphasizing beforehand that $X\neq U$ in the nonlinear solvation scenerios.
The reorganization energy
\be\label{lamd}
 \lambda = \la U \ra_{0} = -\la U \ra_{1} > 0,
\ee
where $\la \hat O \ra_{a} \equiv
{\rm tr} [\hat O\rho^{\E;{\rm eq}}_{a}(T)]$.
\begin{figure}
\includegraphics[width=0.35\textwidth]{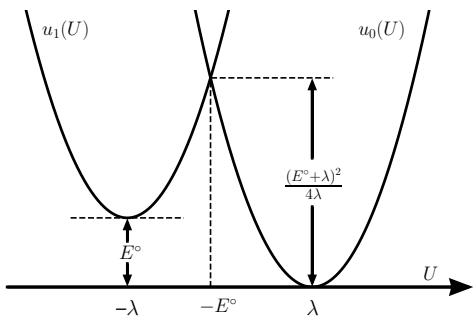}
\caption{The solvation potentials, $u_0(U)$ and $u_1(U)$ of \Eq{u_alpha_U}.
The curve--crossing, $u_0(U^{\ddag})=E^{\circ}+u_1(U^{\ddag})$, occurs
at $U^{\ddag}=-E^{\circ}$. This determines
the activation energy to be the value of $\Delta G^{\ddag} = u_0(U^\ddag)
=(E^{\circ}+\lambda)^2/(4\lambda)$.  
%Here, the reaction coordinate (RC) is $-U$.
}\label{fig1}
\end{figure}
One may elaborate the Marcus' ET theory
with the solvation potentials
$\{u_{\alpha}(U)\}$ that are thermodynamic
measures. Each projects
the microscopic solvent potentials
$\{v_{\alpha}({\bf x})\}$ onto the solvation coordinate as $\{u_{\alpha}(U)\}$, and
$U=u_1(U)-u_0(U)$ is linear and one--dimensional.
%%%
The resultant Boltzmann distribution proportional to
$e^{-\beta u_{a}(U)}$
is a Gaussian function, with
the mean value, \Eq{lamd}, and the variance.
The latter takes a common value as $U$ is linear.
The classical high--temperature
fluctuation--dissipation theorem\cite{Wei12,Yan05187}
gives rise to the Gaussian variance 
\be\label{classical_FDT}
\la\delta U^2\ra_{0} = \la U^2\ra_{0}-\la U\ra^2_{0}
 \approx 2\lambda k_{B}T.
\ee
It together with \Eq{lamd} leads to
\be\label{u_alpha_U}
 u_0(U) = \frac{1}{4\lambda}(U - \lambda)^2
\ \ \,\text{and} \ \ \,
 u_1(U) = \frac{1}{4\lambda}(U + \lambda)^2.
\ee
Figure \ref{fig1} depicts these
solvation potentials. The resultant
activation energy $\Delta G^{\ddag}$ via the curve--crossing condition
agrees perfectly with that in \Eq{ET_rate}.

\subsection{FGR elaboration} \label{thsec2A}
Let us start with the FGR derivations.
The universal FGR rate formula reads
\be\label{fgr}
 k=\frac{2V^2}{\hbar^2}{\rm Re}\!\int_{0}^{\infty}\!\!\!{\rm d}t\,e^{-\frac{i}{\hbar}E^{\circ}t}
%\nl &\quad\times
\Big\la\!\exp_{+}\!\Big[\!-\frac{i}{\hbar}\!\int_{0}^{t}\!{\rm d}\tau\, U(\tau)\Big] \Big\ra.
\ee
%%%
Here, $U(t)\equiv e^{ih_0t/\hbar}Ue^{-ih_0t/\hbar}$ and the
the ensemble average, $\la\,\cdot\,\ra$,
runs over the initial $\rho_0(U)=e^{-\beta u_0(U)}/{\cal Z}$. Here, the partition function
\be\label{Zdef0}
{\cal Z}\equiv  \int_{-\infty}^{\infty}\!\! e^{-\beta u_{0}(U)}\d U=\sqrt{4\pi\lambda k_{B}T}.
\ee
%%%
In the static limit, $U(\tau)\approx U\equiv u_1(X)-u_0(X)$,
one can then perform the time integration first, resulting in \Eq{fgr} the expression,
\be\label{FGR_k}
 k=\frac{2\pi V^2}{\hbar}\La  \delta[u_1(U)-u_0(U)+E^{\circ}]\Ra.
\ee
This together with  $\rho_0(U)=e^{-\beta u_0(U)}/\sqrt{4\pi\lambda k_{B}T}$ recovers the Marcus' ET rate in \Eq{ET_rate}.

\subsection{RRKM analogue} \label{theSec2B}
 Presented above is the FGR elaboration. Now turn to the RRKM analogue to derive \Eq{ET_rate}.
In analogy to the RRKM theory, we apply the ergodicity description.\cite{Berxxxx}
Consider a ET system of total energy $E$ at the donor state $|0\ra$, and therefore
\be\label{Et} 
E=\frac{1}{2}m_{\rm eff}\dot U^2+u_0(U).
\ee
The first term  in \Eq{Et} represents the kinetic energy, where 
$\dot U\equiv \d U/\d t$ is  the moving velocity of the solvation coordinate and
$m_{\rm eff}$ is the effective mass of the solvation collective species to be identified below.
 In the nonadiabatic ET pictures, the reaction rate 
\be \label{kna}
k(E)=\nu(E) P_{0\rightarrow 1}(E).
\ee
Here, $\nu(E)$ is the frequency of the system arriving at the crosspoint $U^{\ddag}$, where the ``activated complex'' is located. It is twice as large as the temporal frequency (angular frequency divided by $2\pi$) reading
\be 
\nu(E)=2\cdot(\w/2\pi),
\ee
where $\w$ is the angular frequency of the ET system at the donor state, and the factor $2$ is due to the fact that the system arrives at the crosspoint twice in each period.
In \Eq{kna}, $P_{0\rightarrow 1}(E)$ is the probability of the electron transferring
from $|0\ra$ to $|1\ra$ 
at the crosspoint of two adiabatic potential surfaces $u_0(U)$ and $u_1(U)$. It adopts the Landau--Zener (LZ) form as \cite{Lan3246,Zen32696,Wit058428}
\be \label{LZ}
P_{0\rightarrow 1}(E)=1-\exp\Bigg[-\frac{2\pi V^2/\hbar}{\big|u'_0(U^\ddag)-u'_1(U^\ddag)\big|\big|\dot U^\ddag\big|}\Bigg].
\ee
In \Eq{LZ},  $u'(U^\ddag)\equiv \d u/\d U|_{U=U^\ddag}$ is the slope of potential surface at the crosspoint, and the velocity $|\dot U^\ddag|=[2(E-\Delta G^{\ddag})/m_{\rm eff}]^{\frac{1}{2}}$ according to \Eq{Et}.
Furthermore, according to the \Fig{fig1}, at the crosspoint $U^{\ddag}=-E^{\circ}$ [\cf\,\Eq{u_alpha_U}],  
\be \label{us4}
|u'_0(U^{\ddag})-u'_1(U^{\ddag})|=1.
\ee  
These give rise to
\be\label{aaa}
P_{0\rightarrow 1}(E)=1-\exp\Bigg[-\frac{2\pi V^2/\hbar}{[2(E-\Delta G^{\ddag})/m_{\rm eff}]^{\frac{1}{2}}}\Bigg].
\ee
In the weak coupling regimes, \Eq{aaa} can be approximated as 
\be \label{us5}
P_{0\rightarrow 1}(E)\simeq \frac{2\pi V^2/\hbar}{[2(E-\Delta G^{\ddag})/m_{\rm eff}]^{\frac{1}{2}}},
\ee
and then we can obtain
\be 
k(E)=\nu(E) P_{0\rightarrow 1}(E)=\frac{(2\w^2m_{\rm eff})^{\frac{1}{2}} V^2/\hbar}{(E-\Delta G^{\ddag})^{\frac{1}{2}}}.
\ee

Similar to the RRKM theory, the overall rate of ET at temperature $T$ reads
\be \label{rrkms}
k=\int_{\Delta G^{\ddag}}^{\infty}\!\!{\rm d}E\, k(E)\frac{e^{-\beta E}}{Z},
\ee
with the partition function
\be 
Z\equiv \int_{0}^{\infty}\!\!{\rm d}E\,e^{-\beta E}=k_{B}T.
\ee
After some simple algebra, we obtain from \Eq{rrkms} that
\be 
k= \frac{(2\w^2m_{\rm eff})^{\frac{1}{2}}V^2/\hbar}{\sqrt{ k_{B}T/\pi}}
  \exp\bigg[-\frac{(E^{\circ}+\lambda)^2}{4\lambda k_{B}T}\bigg].
\ee
Compare with the Marcus' rate in \Eq{ET_rate}, we know that 
\be \label{rec}
2\w^2m_{\rm eff}=1/\lambda.
\ee
This fact is also verified via the detailed analysis of the quadratic solvation scenarios in \Sec{thsec4A} [\cf\,\Eq{meff} with $\theta=1$]. We then recover the Marcus' nonadiabatic ET rate formula,
\Eq{ET_rate}, via this RRKM analogue.

 In the following section, according to this RRKM analogue, we will construct
a generalized rate formula  for ET
processes in the nonlinear solvation scenarios, where the
multiple curve--crossing of solvation potentials exists.

\section{Nonlinear solvation scenario}

In general, the  energy difference, $U=h_1-h_0$, defined in \Eq{U_def} is nonlinear, i.e., $X\neq U$, 
with $X$ up to the linear part of $U$. Therefore, the thermodynamic solvation potentials hereafter
are $u_0(X)$ and $u_1(X)$, instead of $u_0(U)$ and $u_1(U)$, associated with the reactant and product, respectively. 
In the nonlinear solvation scenarios, 
there may exist multiple curve--crossing of solvation potentials, resulting in multiple crosspoints.
Let $X^{\ddag}_n$ be the $n$th crossing point,
with the barrier height,  $\Delta G^{\ddag}_n$,
satisfying
\be\label{DelG_def}
 \Delta G^{\ddag}_n = u_0(X^{\ddag}_n) = u_1(X^{\ddag}_n) + E^{\circ}.
\ee
%%%
In the remaining part of this section, we elaborate the ET rate formula via both the FGR and the RRKM analogue. 

\subsection{FGR elaboration} \label{thesec3A}
Similar to \Eq{FGR_k}, the FGR rate in this scenario reads
\be\label{FGR_k2}
 k=\frac{2\pi V^2}{\hbar}\La  \delta[u_1(X)-u_0(X)+E^{\circ}]\Ra
\ee
with 
the ensemble average, $\la\,\cdot\,\ra$,
runs over the initial state $\rho_0(X)=e^{-\beta u_0(X)}/{\cal Z}$ where
\be\label{Zdef}
{\cal Z}\equiv  \int_{-\infty}^{\infty}\!\! e^{-\beta u_{0}(X)}\d X.
\ee
 Mathematically, the evaluation 
is carried out with the roots of involving
delta function by solving $u_1(X)-u_0(X)+E^{\circ}=0$.
The solutions just satisfy the second expression of \Eq{DelG_def}.
Now by applying the identity,
\be\label{del_formula}
\int_{-\infty}^{\infty}\!\!{\rm d}x\,f(x)\delta[g(x)]=\sum_{i}\frac{f(x_i)}{|g'(x_i)|},
\ee
where $g'(x_i)$ denotes the slope of $g(x)$ at the root $x=x_i$. Note that $f(x_i)=0$ if $x_i$ is not a real root
and $f(x)$ is a real distribution function,
such as $\rho_0(X)$ in the present study. The resultant rate  reads
\be \label{kfgr2}
k=\frac{2\pi V^2/\hbar}{{\cal Z}}\sum_n\! \frac{1}{\big|u'_0(X_n^\ddag)\!-\!u'_1(X_n^\ddag)\big|}\!\exp\bigg(\!-\frac{\Delta G^{\ddag}_n}{ k_{B}T}\bigg).
\ee
There are two limiting scenarios, as follows.
One is the Marcus' linear solvation scenario,
with \Eq{u_alpha_U} being just
$u_0(X)=(X-\lambda)^2/(4\lambda)$
and $u_1(X)=(X+\lambda)^2/(4\lambda)$.
In this case, $|u'_0(X)-u'_1(X)|=1$.
It is easy to verify that all the results
in \Sec{thsec2} are reproduced.  Another limiting scenario is $u'_0(X^{\ddag})=u'_1(X^{\ddag})$.
This is the case that two potential curves are just barely touched. In this critical  case, \Eq{kfgr2} breaks down due to the divergence. From the perception of RRKM analogue elaborated in \Sec{theSec2B}, we know that it is invalid to do the approximation in \Eq{us5} when $|u'_0(X)-u'_1(X)|$ is relatively small. It leads to the divergence in this limiting case.

\subsection{RRKM analogue}\label{thesec3B}
Due to the breakdown of approximation in \Eq{us5}, we directly apply the original LZ form of transition probability.
Similarly, consider a ET system of total energy $E$ at the donor state $|0\ra$, and therefore
\be
E=\frac{1}{2}m_{\rm eff}\dot X^2+u_0(X).
\ee
The reaction rate associated with $E$ through $n$th crosspoint is then
\be
k_n(E)= \nu_n(E) P_{0\rightarrow 1;n}(E).
\ee
Similarly,
\be 
\nu_n(E)=2\cdot(\w/2\pi),
\ee
and the LZ form of transition probability reads
\be \label{LZ2}
P_{0\rightarrow 1;n}(E)=1-\exp\Bigg[-\frac{2\pi V^2/\hbar}{\big|u'_0(X^\ddag_n)-u'_1(X^\ddag_n)\big|\big|\dot X_n^\ddag\big|}\Bigg],
\ee
with $\big|\dot X_n^\ddag\big|=[2(E-\Delta G_n^{\ddag})/m_{\rm eff}]^{\frac{1}{2}}$.
Then the overall rate of ET at temperature $T$ reads
\be 
k=\sum_n \int_{\Delta G^{\ddag}_n}^{\infty}\!\!{\rm d}E\, k_n(E)\frac{e^{-\beta E}}{Z},
\ee
with
$
Z=k_{B}T.
$
Here, the sum runs over all the reactive channels
to the product.
We then obtain
\begin{align}
k= \frac{\w}{\pi k_B T}\sum_n \int_{\Delta G^{\ddag}_n}^{\infty}\!\!{\rm d}E\, P_{0\rightarrow 1;n}(E)e^{-\beta E},
\end{align}
with
\be 
P_{0\rightarrow 1;n}(E)\!=\!1-\exp\!\!\Bigg[\!-\frac{(2m_{\rm eff})^{\frac{1}{2}}\pi V^2/\hbar}{\big|u'_0(X^\ddag_n)\!-\!u'_1(X^\ddag_n)\big|(E-\Delta G_n^{\ddag})^{\frac{1}{2}}}\!\Bigg].
\ee
Use the formula
\be 
\int_{0}^{\infty}\!{\rm d}t\,te^{-2a/t-bt^2}=\frac{a^2}{2\sqrt{\pi}}G^{3,0}_{0,0}(a^2b|_{-1,-1/2,0}),
\ee
where $G^{3,0}_{0,0}(x|_{-1,-1/2,0})$ is the corresponding Meijer G-function, and we can obtain
\be \label{rate_new}
k=\frac{\w}{\pi}\sum_n \eta_n \exp\bigg(-\frac{\Delta G^{\ddag}_n}{ k_{B}T}\bigg),
\ee
with
\be 
\eta_n=1-\frac{\kappa_n^2G^{3,0}_{0,0}(\beta\kappa_n^2|_{-1,-1/2,0})}{\sqrt{\pi}k_B T}
\ee
where
\be 
\kappa_n=\frac{(m_{\rm eff}/2)^{\frac{1}{2}}\pi V^2/\hbar}{\big|u'_0(X^\ddag_n)-u'_1(X^\ddag_n)\big|}.
\ee
Now consider the limiting scenario where $u'_0(X^{\ddag})=u'_1(X^{\ddag})$. We have
$
\eta_n\rightarrow 1
$,
and the corresponding rate formula in \Eq{rate_new} does not suffer the divergence any more. In this rate formula, the angular frequency $\w$ and the effective mass $m_{\rm eff}$ are to be determined with respect to concrete settings. In the next section, as an illustrative example, we will consider the
quadratic solvation scenarios,
on the basis of physically
well--supported descriptors.\cite{Xu17395,Xu18114103,Liu18245}

\section{Electron transfer in quadratic solvation environments}
\label{thsec4}

\subsection{Quadratic thermodynamic potentials analysis}
\label{thsec4A}
It is worth emphasizing that a physical
description on the system--environment couplings in solutions
should satisfy the invariance requirement, with respect
to the reference environment. For a linear coupling solvation environment,
this requirement is automatically satisfied.
%%%
However, it is a nontrivial task even
for quadratic solvation environments.\cite{Xu17395,Xu18114103,Liu18245}
Presented below is the recently developed
quadratic solvation descriptors scheme\cite{Xu18114103}
that satisfies the aforementioned reference--environment
invariance requirement.

Recall that $h_{0}$ and  $h_{1}$ are the environment Hamiltonians associated with the electronic system in donor and acceptor states, respectively [\cf\Eq{HT0}].
In contact to the solvation modes description, we express these two environment Hamiltonian in the Calderia--Leggett's form,\cite{Cal83587}
\be\label{hB_solvA}
\begin{split}
  h_{0} &= \frac{\w}{2}(\hat p^2+\hat x^2)
    +\!\sum_{k}\!\frac{\ti\w_k }{2}\Big[
     \ti p^2_k+\!\Big(\ti{x}_k\!-\!\frac{\ti c_k}{\ti\w_k}\hat x \Big)^2
    \Big],\\
  h_1 &=\!\frac{\w'}{2}(\hat p^{\prime 2}+\hat x^{\prime 2})
    +\!\sum_{k}\!\frac{\ti\w'_k}{2} \!\Big[
     \ti p^{\prime 2}_k+\!\Big(\ti{x}'_k\!-\!\frac{\ti c'_k}{\ti\w'_k}\hat x'\Big)^2
    \Big].
\end{split}
\ee
Physically, each expression describes a Brownian oscillator (first term), under the influence of secondary environment.
One can readily obtain the generalized Langevin equation, in which the friction kernel reads \cite{Yan05187}
\be\label{zeta_t}
\begin{split}
  \zeta(t)&=\w\sum_{k} \frac{\ti c^2_k}{\ti\w_k}\cos(\ti\w_k t),
  \\
  \zeta'(t)&=\w'\sum_{k} \frac{\ti c^{\prime 2}_k}{\ti\w_k'}\cos(\ti\w'_k t).
\end{split}
\ee
The $h_{1}$-environment is subject to linear--displacements and frequency--shifts in relation to $h_{0}$. This can be described as
\bsube\label{xB_all}
\begin{align}
 \hat x' &= (\w'/\w)^{\frac{1}{2}}(\hat x-d)
   \equiv \theta^{\frac{1}{2}}(\hat x-d),
\label{xB_a}\\
 \ti{x}'_k &= (\ti\w'_{k}/\ti\w_{k})^{\frac{1}{2}}(\ti{x}_k-\ti d_{k})
   \equiv \ti\theta^{\frac{1}{2}}_{k}(\ti{x}_k-\ti d_{k}).
\label{xB_b}
\end{align}
\esube
The kinetics energies in $h_{0}$ and $h_{1}$ are the same.
Similar to \Eq{U_def}, the solvation energy with respect to each individual reference environment is given respectively by
\bsube\label{del_hB_def_all}
\begin{align}
 \delta h_{0} \equiv h_{1}-h_{0}
 &= \alpha_0 + \alpha_1\hat x+\alpha_2\hat x^2,
\label{del_hB_a}\\
 \delta h_{1} \equiv h_{0}-h_{1}
 &= \alpha'_0 + \alpha'_1\hat x'+\alpha'_2\hat x^{\prime 2}.
\label{del_hB_b}
\end{align}
\esube
The involving descriptors $\{\alpha_i\}$ and $\{\alpha'_i\}$
should satisfy
the reference--environment invariance that reads
\begin{align}\label{HT_HT}
&\quad\,h_0+\hat Q_{\tS}(\alpha_0+\alpha_1\hat x+\alpha_2\hat x^2)
\nl&
 = h_1+(1-\hat Q_{\tS})
  (\alpha'_0+\alpha'_1\hat x'+\alpha'_2\hat x^{\prime 2}).
\end{align}
Here $\hat Q_{\tS}$ describes the dissipative system mode on which the environment acts.
In contact with the ET system, \Eq{HT0}, $\hat Q_{\tS}=|1\ra\la 1|$.
To determine the descriptors, we adopt the
linear--displacement--mapping method,
which results in\cite{Xu18114103}
\be \label{lam_def}
\lambda\equiv \la \delta h \ra_{\alpha_2=0}=\frac{1}{2}\w d^2,
\ee
and
\be\label{alpha_all}
\begin{split}
\!\!\alpha_0\! = \! \lambda\theta^{2},
\quad
 \alpha_1\!=\!-(2\lambda\w)^{\frac{1}{2}}\theta^{2},
\quad
 \alpha_2 \!= \!\frac{\w}{2}(\theta^{2}-1),
\end{split}
\ee
whereas
\be\label{alphap_all}
\begin{split}
\!\!\alpha'_0\! = \! \lambda ,
\quad
 \alpha'_1 \!= \!(2\lambda\w/\theta)^{\frac{1}{2}},
 \quad
 \alpha'_2\! = \!\frac{\w}{2\theta}(1-\theta^2).
\end{split}
\ee
The derivations were made in the constraint
of the environment--reference invariance, \Eq{HT_HT}.
%%%
Moreover, the linear displacement mapping
ansatz implies also\cite{Xu18114103}
\be
\zeta(t)=\zeta'(t).
\ee

 To obtain the thermodynamic potentials,
let us start with $v_{0}(\hat{\bf x})$ and $v_{1}(\hat{\bf x}')$,
the potential energies of $h_0$ and $h_1$, respectively [\cf\Eq{hB_solvA}].
Project them to the same solvation coordinate $\hat x$, resulting in the thermodynamic potentials, $u_{0}(\hat x)$ and $u_{1}(\hat x)$. By doing this, we can recast \Eq{del_hB_a} as
\be \label{del_u}
U=u_{1}(\hat x)-u_{0}(\hat x)= \alpha_0 + \alpha_1\hat x+\alpha_2\hat x^2.
\ee
To facilitate the analysis in line with \Eq{u_alpha_U} and \Fig{fig1}, we introduce
\bsube \label{tiU_def}
\begin{align}
X&\equiv \alpha_{0}+\alpha_{1}\hat x,
\label{tiU_a}\\
X'&\equiv \alpha'_{0}+\alpha'_{1}\hat x'.
\label{tiU_b}
\end{align}
\esube
The thermodynamic potentials associated with $h_0$ and $h_1$ would read
\bsube\label{uB}
\begin{align}
u_{0}(X)=&\frac{1}{4\Theta_0\lambda_0}(X-\lambda_0)^2,
\label{uB_a}\\
u_{1}(X')=&\frac{1}{4\Theta_1\lambda_1}(X'-\lambda_1)^2.
\label{uB_b}
\end{align}
\esube
The parameters $\lambda_0$, $\lambda_1$, $\Theta_0$ and $\Theta_1$ are to be determined.
Firstly, the definitions in \Eq{tiU_def} tells us
\be
\la X\ra_0=\alpha_0\ \ \ \ \text{and}\ \ \ \ \la X'\ra_{1}=\alpha_0',
\ee
and therefore,
\be
\lambda_0=\alpha_0=\lambda\theta^2\ \ \ \ \text{and}\ \ \ \ \lambda_1=\alpha_0'=\lambda.
\ee
To proceed, we need to relate the primed quantities to the unprimed counterparts. By using \Eqs{xB_a}, (\ref{lam_def}) and (\ref{tiU_a}), we obtain
\be
X'=[\alpha'_{0}-\alpha'_{1}(2\lambda\theta/\w)^{\frac{1}{2}}]+\alpha'_{1}\theta^{\frac{1}{2}}(X-\alpha_0)/\alpha_1.
\ee
Here, $\alpha'_0=\theta^{-2}\alpha_0$, $\alpha'_1=-\theta^{-5/2}\alpha_1$ and $\alpha_0/\alpha_1=-d/2$, as inferred from \Eq{alphap_all} versus \Eq{alpha_all}.  Therefore,
$
X'=-X/\theta^{2}
$,
which leads to \Eq{uB} the expression,
\bsube\label{uuuu}
\begin{align}
u_0(X)=&\frac{1}{4\Theta_0\lambda\theta^2}(X-\lambda\theta^2)^2,
\\
u_1(X)=&\frac{1}{4\Theta_1\lambda\theta^4}(X+\lambda\theta^2)^2.
\end{align}
\esube
Together with \Eq{tiU_a}, we obtain %(\ref{uB_a}) and (\ref{upU}),
\begin{align}%\label{del_u_2}
\ u_1-u_0
&=\frac{(\alpha_{0}+\alpha_{1}\hat x+\lambda\theta^2)^2}
       {4\Theta_1\lambda\theta^4}
\nl&\quad
 -\frac{(\alpha_{0}+\alpha_{1}\hat x-\lambda\theta^2)^2}
       {4\Theta_0\lambda\theta^2}.
\end{align}
By comparing with \Eq{del_u}, we have also
\begin{align}\label{tobesolved}
\Theta_0=\Theta_1^{-1}=\theta^{2},
\end{align}
which further leads to \Eq{uuuu} the expression,
\bsube\label{UUU}
\begin{align}\label{UUU0}
u_0(X)=&\frac{1}{4\lambda\theta^4}(X-\lambda\theta^2)^2,
\\\label{UUU1}
u_1(X)=&\frac{1}{4\lambda\theta^2}(X+\lambda\theta^2)^2.
\end{align}
\esube
In \Eq{UUU} are the thermodynamic solvation potentials to be used later.

%It is worth noting that there is another common choice of the reaction coordinate in literatures. That is $U$ in \Eq{del_u}, rather then $X$ in \Eq{tiU_a}. This choice will lead to the anharmonicity of the solvation potentials.\cite{Sma037470}

Now turn to the kinetic energy term to obtain the effective mass $m_{\rm eff}$. In $h_0$, according to \Eqs{hB_solvA} and (\ref{tiU_a}), we can recast the kinetic energy term as
\be 
\frac{1}{2}\w\hat p^2=\frac{1}{2}\w\alpha_1^2 P^2,
\ee
where $[\hat x,\hat p]=[X,P]=i$. Therefore, the canonical equation of motion gives
\be 
\dot X=\w\alpha_1^2 P.
\ee
The kinetic energy term can be then expressed as
\be 
\frac{1}{2}\w\alpha_1^2 P^2=\frac{1}{2}m_{\rm eff} \dot X^2,
\ee
with 
\be \label{meff}
m_{\rm eff}=\frac{1}{\w\alpha_1^2}=\frac{1}{2\lambda\w^2\theta^4}.
\ee
This recovers \Eq{rec} in \Sec{theSec2B} when $\theta=1$.

\subsection{ET rate in quadratic solvation environments}\label{thsec4B}

\begin{figure}
\includegraphics[width=0.48\textwidth]{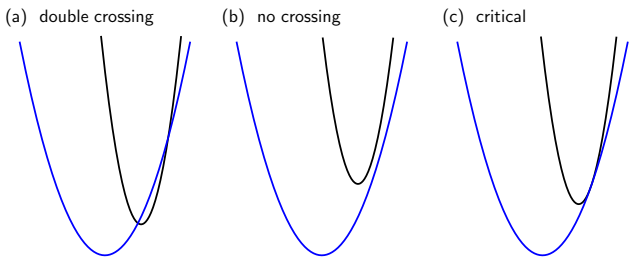}
\caption{Three representing
scenarios, in relation to
the curve--crossing behaviors, involving
two solvation potentials with different curvatures ($\theta\neq 1$).
(a) 	Double crossing case;
(b) No crossing case;
(c) Critical case:  $u'_0(X^{\ddag})=u'_1(X^{\ddag})$.
The potential surfaces are plotted as the functions of $X$ [\cf \Eq{UUU}], and $X\neq U$ in this quadratic solvation scenarios [\cf \Eqs{del_u} and (\ref{tiU_def})].}
\label{fig2}
\end{figure}

In \Sec{thsec4A}, we retain \Eq{U_def}
as the general definition of $U$,
and \Eq{lamd} for $\lambda$ is no longer valid, whenever
$\theta\neq 1$.
Note also that $\theta >0$, as implied in \Eq{alphap_all}. Apparently, \Eq{UUU0} leads to \Eq{Zdef} the value
\be\label{Z_value}
{\cal Z}=(4\pi\lambda k_{B}T)^{\frac{1}{2}}\theta^2.
\ee
According to \Eq{UUU}, shown in \Fig{fig2} are three representing
scenarios, in relation to
the curve--crossing behaviors, involving
two solvation potentials with different curvatures ($\theta\neq 1$).
%%%
As discussed earlier, \Fig{fig2}(a) represents the multiple curve-crossing case, with
the ET rate $k>0$, whereas (b) goes with
$k=0$, as there are no crossing points.
Figure \ref{fig2}(c) represents the critical scenario of $u'_0(X^{\ddag})=u'_1(X^{\ddag})$.

As seen below, one can define the quadratic characteristic parameter,
\be\label{qc_def}
 q_c\equiv \frac{(1-\theta^2)E^{\circ}}{\theta^2\lambda}.
\ee
The three representing scenarios of \Fig{fig2}
are related to (a) $1+q_c >0$, (b) $<0$
and (c) $=0$, respectively.

 As the kinetic rate process is concerned,
in the following we focus only on the case of \Fig{fig2}(a).
For the specified solvation potentials, \Eq{UUU},
we obtain \Eq{DelG_def} the solutions,
\be\label{xpm}
	X^{\ddag}_{\pm} = -E^{\circ}+\Lambda_{\pm},
\ee
with [\cf\Eq{qc_def}]
\be \label{lampm}
\Lambda_{\pm}= \frac{\lambda\theta^2}{1-\theta^2}(\sqrt{|1+q_c|}\pm\theta)^2.
\ee
%%%
The corresponding activation energy reads
\begin{align}\label{gpm}
\Delta G_{\pm}^{\ddag}=u_0(X_{\pm})
=\frac{1}{4\lambda\theta^4}(E^{\circ}
+\lambda\theta^2- \Lambda_{\pm})^2.
\end{align}
 The resultant FGR rate acquires the form of \Eq{kfgr2}, which reads
\be
k=\frac{2\pi V^2/\hbar}{{\cal Z}}\eta^\text{\tiny FGR}\sum_{\pm}\!\exp\bigg(-\frac{\Delta G^{\ddag}_{\pm}}{ k_{B}T}\bigg).
\ee
with
\be\label{eta_FGR}
 \eta^\text{\tiny FGR}
 %(X^{\ddag}_n) 
 = \frac{\theta}{\sqrt{|1+q_c|}}.
\ee
In the
linear solvation case ($\theta=1$), 
$q_c=0$ via \Eq{qc_def}, resulting in
the value of $1$ for  \Eq{eta_FGR}.
%%%
However, when  $1+q_c\rightarrow 0$,
$\eta^\text{\tiny FGR}\rightarrow \infty$, suffering a divergence.
This is the scenario depicted in \Fig{fig2}(c),
where the slopes at the pseudo-crossing point
are of $u'_0(X^\ddag)=u'_1(X^\ddag)$.  
In this limiting case, we must consult the RRKM analogue.
{\color{black}
 \Eq{rate_new} tells us 
\be 
k=\frac{\w}{\pi}\exp\bigg[-\frac{\lambda\theta^4}{ k_{B}T(1-\theta^2)^2}\bigg]
\ee 
in this limiting case, where $\w$ is as defined in \Eq{hB_solvA}.
}

\section{Concluding remarks}
\label{thsum}

  In summary, we have proposed a RRKM analogue to ET processes.
Not only does it recover the original Marcus' rate,
the proposed theory is also applicable to the nonlinear solvation scenarios, where
multiple curve--crossing of solvation potentials exists.
%%%
It is noticed that the original RRKM theory is
concerned only with adiabatic gas--phase reactions.%
\cite{Ric271617,Ric28617,Kas28225,Kas281065,Mar51894,Mar52359,Hol96}
We elaborate this widely used kinetic mechanism
from the analogous aspect, as detailed in \Sec{theSec2B}
and \Sec{thesec3B}. 
The obtained RRKM analogue is constructed
on the basis of the ergodicity description, where we have also used the LZ formula for the transition probability.

 We also revisit the corresponding FRG formula, with some critical comments against
the RRKM analogue proposed in this work.
Both approaches result in the Arrhenius--type 
expression of rate, with same activation energies
in exponentials, but distinct pre-exponential
coefficients. We scrutinize their differences
particularly in the scenario of \Fig{fig2}(c),
as highlighted at the end of \Sec{thsec4B}.
%%%
While the new theory gives  the rate a finite value in this scenario,
the FGR would result in $k\rightarrow\infty$.
%%%
This also highlights the critical importance of
an appropriate pre-exponential factor,
in particular when the curve--crossing is degenerate.

However, we notice there still exists a discontinuity of the ET rate between the critical case ($k\neq 0$) and the no crossing case ($k=0$). To tackle with this discontinuity, we may further consult the Rosen--Zener nonadiabatic transition probability for the no crossing cases.
As introduced in the literature, \cite{Zhu964159} we may adopt
\be \label{zhueq}
P_{0\rightarrow 1}(E)=1-\frac{\sinh[(D^2-1)\Gamma]}{\sinh(D^2\Gamma)}e^{-\Gamma}
\ee
in place of \Eq{LZ2}. In \Eq{zhueq},
\be 
\Gamma=\frac{2\pi V^2/\hbar}{\big|u'_0(X^\ddag)-u'_1(X^\ddag)\big|\big|\dot X^\ddag\big|},
\ee
and
%\be 
%D=\sqrt{1+\frac{4V^2}{[u_0(X^\ddag)-u_1(X^\ddag)-E^{\circ}]^2}}.
%\ee
\be 
D=\bigg\{1+\frac{4V^2}{[u_0(X^\ddag)-u_1(X^\ddag)-E^{\circ}]^2}\bigg\}^{\frac{1}{2}}.
\ee
When there is no crossing, $X^\ddag$ denotes the real part of the complex solutions of the equation $u_0(X) = u_1(X) + E^{\circ}$, which leads to $u_0(X^\ddag)-u_1(X^\ddag)-E^{\circ} \neq 0$. However, in the crossing cases, $u_0(X^\ddag)-u_1(X^\ddag)-E^{\circ}=0$, which results in $D\rightarrow \infty$ and therefore $\sinh[(D^2-1)\Gamma]/\sinh(D^2\Gamma)\rightarrow 1$. This gives back to the LZ formula, and all results in \Sec{thesec3B} are reproduced. Consequently, this modification would eliminate the existing discontinuity of the ET rate between the critical case and the no crossing case. It not only revisits the RRKM analogue results in the crossing and critical cases, but also leads to $k\neq 0$ in the no crossing cases.

{\color{black}It is also worth noting that the present work
treats the solvent environments in the static
and classical limit.
%%%
The failure of FGR treatment
in the critical case [\cf\,\Fig{fig2}(c)]
may imply that the dynamical fluctuation
is an intrinsic nature associated
with the nonlinearity.
 Furthermore, in the quantum regime, the nonlinear effects may become more prominent.\cite{Xu17395,Xu18114103,Liu18245} The chemical kinetics based on microscopic descriptions, together with  exact quantum dissipative dynamics, may help further understanding the dynamical nonlinear effects in the ET rate processes in this regime.
 }

%\subsection*{Data Availability}
%The data that support the findings of this study are available from the corresponding author upon reasonable request.

\begin{acknowledgments}
 Support from
the Ministry of Science and Technology of China (Nos.\ 2017YFA0204904 \&
2016YFA0400904),
the Natural Science Foundation of China (Nos.\ 21633006)
and Anhui Initiative in Quantum Information Technologies
is gratefully acknowledged.
\end{acknowledgments}

%\bibliographystyle{../aip}
%\bibliography{../bibrefs}

\end{document}